\documentclass[aps,pre,twocolumn,superscriptaddress]{revtex4-2}

\usepackage{amsmath,amssymb,amsfonts}
\usepackage{graphicx}
\usepackage{hyperref}
\usepackage{appendix}
\usepackage{nicefrac}
\usepackage{color}
\usepackage{float}
\usepackage[normalem]{ulem}

\begin{document}
	\title{Impact of Higher-Order Interactions on Collective Motion}
	\author{Maryam Masoumi}
	\affiliation{Department of Physics, Shahid Beheshti University, Evin, Tehran 1983969411, Iran}
	\author{Amir Kargaran}
	\email{amir.kargaran@ipm.ir}
	\affiliation{School of Biological Sciences, Institute for Research in Fundamental Sciences (IPM), 193955746, Tehran, Iran}
	\author{Reza Jafari}
	\email{g\_jafari@sbu.ac.ir}
	\affiliation{Department of Physics, Shahid Beheshti University, Evin, Tehran 1983969411, Iran}
	\date{}
	\begin{abstract}
        Collective motion in self-propelled particle systems has been widely studied using the Vicsek model, which relies on pairwise alignment interactions. We introduce a generalized Vicsek model that incorporates higher-order (triadic) alignment interactions. Using agent-based simulations and mean-field theory, we demonstrate that pure triadic alignment induces a discontinuous phase transition, evidenced by hysteresis, a double-well free-energy landscape, and a Binder cumulant minimum that deepens with system size, whereas the standard pairwise model exhibits a continuous transition at the same system sizes. We further show that higher-order interactions require higher particle densities to sustain collective order and produce sharper fluctuation peaks near the transition with lower critical noise. These results establish that the microscopic structure of the alignment interaction, whether pairwise or many-body, is an independent control parameter for the order of the phase transition in active matter, with implications for understanding collective behavior in biological and synthetic systems.
	\end{abstract}
	\maketitle
    
	\section{Introduction}
    Collective motion, the spontaneous emergence of coordinated behavior in interacting agents, is ubiquitous across many systems, including bird flocks, bacterial colonies, and cellular systems~\cite{Sumpter2006,Couzin2003,VicsekBactery1996,MehesVicsek2014}, as well as human crowds, robotic swarms, and financial markets~\cite{Warren2018,Castellano2009,Afsharizand2020,SAEEDIAN2019}. It also appears in a wide range of active and physical matter systems~\cite{Marchetti2013,VICSEK2012}. The Vicsek model~\cite{VicsekPRL1995,Ginelli2016} provides the minimal framework for this phenomenon: self-propelled particles moving at constant speed align with their neighbors under stochastic noise, producing a non-equilibrium transition from disordered motion to collective order. While noise amplitude and particle density are well-established control parameters for this transition, whether the microscopic structure of the alignment interaction itself can independently control the order of the transition has remained an open question. More fundamentally, can collective behavior arise from genuinely higher-order (triadic) interactions, and if so, what are its distinguishing properties?

    The Vicsek framework has been extended in numerous directions, biologically motivated variants, hydrodynamic descriptions, and alternative interaction rules~\cite{AlbanoPRL1996,CzirokPHYA1997,HuepePRL2004,TonerTu1995,TonerTu1998,ChatePRE2008}, as well as models revealing nematic order, cohesion, and coherent structures such as vortices and clusters~\cite{ChatePRL2006,GREGOIRE2003157,DOrsognaPRL2006,lardent2026}. More recent work incorporates environmental refinements, data-driven approaches, and delay effects~\cite{VahabliNC2023,PakpourPhysicsA2024,FalkPRR2021,CelaniPRE2020,Loffler2023}, underscoring the framework's versatility. Despite this breadth, these studies have relied almost exclusively on pairwise alignment, leaving the role of higher-order interactions in active matter largely unexplored.

    Higher-order interactions, which couple three or more agents simultaneously and cannot be decomposed into independent pairs, are now recognized as fundamental in complex systems~\cite{Iacopini2019,Benson2016,Battiston2021,BOCCALETTI2023,TabarCP2025,TabarPRX2024}. In oscillatory and networked systems, they generate abrupt transitions and multistability absent in pairwise models~\cite{Chatterjee2019,Skardal2019,Anwar2024,Zhang2021,leon2025}. In active matter, however, the alignment rule is the dynamical core, and its replacement by a genuinely triadic interaction raises a distinct question: does interaction structure alone, independent of noise or density, control the order of the phase transition?
    
    Here we introduce a minimal generalization of the Vicsek model in which pairwise alignment is replaced or mixed with a genuinely triadic interaction, controlled by a single parameter $\alpha$. Using agent-based simulations and mean-field analysis, we show that higher-order alignment qualitatively alters the transition character, driving discontinuous behavior at system sizes where the pairwise model appears continuous. This provides a direct demonstration that interaction structure is an independent control parameter for the order of the phase transition in active matter.

    \begin{figure}[t]
        \centering
        \includegraphics[width=1\linewidth]{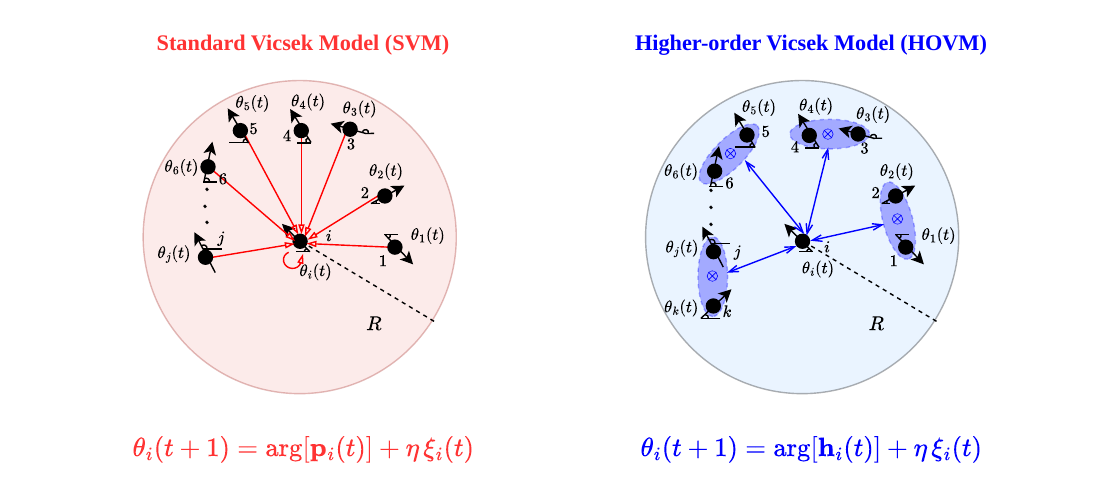}
        \caption{Schematic of alignment interaction rules. Left: Standard Vicsek Model (SVM), in which each particle aligns with the mean velocity of all neighbors within radius $R$. Right: Higher-Order Vicsek Model (HOVM), in which alignment is determined by triadic interactions, each focal particle interacts simultaneously with pairs of neighbors, as defined in Eq.~\eqref{eq:triadic_field}.}
        \label{fig:main-fig1}
    \end{figure}
    
	\section{Model}\label{sec:model}
    The Vicsek model~\cite{VicsekPRL1995} describes $N$ self-propelled 
    particles moving in discrete time with synchronous updates in a 
    two-dimensional square domain of size $L$ with periodic boundary 
    conditions. Each particle $i$ has position $\mathbf{x}_i(t)$, 
    orientation $\theta_i(t)$, and velocity 
    $\mathbf{v}_i(t)=v_0(\cos\theta_i(t),\sin\theta_i(t))$ with constant 
    speed $v_0$. Positions are updated as 
    $\mathbf{x}_i(t+1)=\mathbf{x}_i(t)+\mathbf{v}_i(t)$. Orientations are 
    updated by aligning with the local polarization 
    $\mathbf{p}_i(t)=(v_0|\mathcal{N}_i(t)|)^{-1}
    \sum_{j\in\mathcal{N}_i(t)}\mathbf{v}_j(t)$, where $\mathcal{N}_i(t)$ 
    is the set of particles within metric radius $R$ of particle $i$ 
    (including $i$ itself) and $|\mathcal{N}_i(t)|$ is its size 
    [Fig.~\ref{fig:main-fig1}, left]. The updated orientation is 
    $\theta_i(t+1)=\arg[\mathbf{p}_i(t)]+\eta\,\xi_i(t)$, where $\eta$ is 
    the noise amplitude and $\xi_i(t)$ is uniformly distributed in 
    $[-1/2,1/2]$. Collective motion is quantified by the global polarization 
    $v_a=(Nv_0)^{-1}\left|\sum_i \mathbf{v}_i\right|$, which ranges from 
    zero in the disordered state to one in a fully aligned system.
    
    We extend this framework by introducing higher-order (triadic) 
    interactions [Fig.~\ref{fig:main-fig1}, right]. The interaction is 
    constructed from the symmetric tensor product of two neighbor velocities 
    contracted with the focal particle velocity, generating terms of the form 
    $\big[(\mathbf{v}_j\otimes\mathbf{v}_k)+
    (\mathbf{v}_k\otimes\mathbf{v}_j)\big]\mathbf{v}_i$, which couple 
    triplets of particles and depend explicitly on their relative 
    orientations. For each particle $i$, we define the set of unordered 
    neighbor pairs $\mathcal{T}_i(t)=\{(j,k)\,|\, j,k\in\mathcal{N}_i(t),
    \, j<k\}$ and the higher-order alignment field
    
    \begin{equation}
        \mathbf{h}_i(t)=\frac{1}{v_0^3|\mathcal{T}_i(t)|}
        \sum_{(j,k)\in\mathcal{T}_i(t)}
        \left[(\mathbf{v}_j\cdot\mathbf{v}_i)\mathbf{v}_k
        +(\mathbf{v}_k\cdot\mathbf{v}_i)\mathbf{v}_j\right],
    \label{eq:triadic_field}
    \end{equation}
    
    where $|\mathcal{T}_i(t)|$ is the number of neighbor pairs; if 
    $|\mathcal{T}_i(t)|=0$, we set $\mathbf{h}_i(t)=\mathbf{0}$. This 
    interaction is intrinsically non-pairwise: each term in the sum involves 
    three particles simultaneously and cannot be decomposed into a sum of 
    pairwise terms. The fields $\mathbf{h}_i$ and $\mathbf{p}_i$ are 
    normalized consistently: $\mathbf{p}_i$ is divided by $v_0|\mathcal{N}_i|$, 
    accounting for one power of $v_0$ per velocity and the number of 
    neighbors, while $\mathbf{h}_i$ is divided by $v_0^3|\mathcal{T}_i|$, 
    accounting for three powers of $v_0$ per triplet and the number of 
    neighbor pairs. This ensures $\mathbf{h}_i$ and $\mathbf{p}_i$ remain 
    comparable in magnitude across different local densities.
    
    In the hybrid model, pairwise and higher-order interactions are combined 
    through a mixing parameter $\alpha\in[0,1]$. The effective alignment 
    field is $\mathbf{u}_i(t)=\alpha\,\mathbf{p}_i(t)+(1-\alpha)\,
    \mathbf{h}_i(t)$, and the orientation update rule becomes
    
    \begin{equation}
        \theta_i(t+1)=\arg[\mathbf{u}_i(t)]+\eta\,\xi_i(t).
    \end{equation}
    
    This construction preserves the rotational symmetry of the Vicsek model 
    while introducing a nonlinear dependence on the local polarization. The 
    limits $\alpha=1$ and $\alpha=0$ recover the standard Vicsek model 
    (SVM) and the purely higher-order Vicsek model (HOVM), respectively.

    \section{Simulations}\label{sec:simulations}
    We simulate the hybrid model (Sec.~\ref{sec:model}) in a two-dimensional square domain with periodic boundaries, with particle numbers $N=490$ to $9\times10^3$ and densities $\rho=N/L^2\in[0.1,10]$. Particles move at speed $v_0=0.03$ and interact within radius $R=1$ according to the original Vicsek protocol~\cite{VicsekPRL1995}. For steady-state observables, each noise value $\eta$ is initialized with random orientations, equilibrated for $t_{\rm eq}\ge10\tau_{\max}$, and sampled over $t_{\rm s}\ge5\tau_{\max}$, where $\tau_{\max}$ is the maximal integrated autocorrelation time (see Supplemental Material); averages are taken over $10^3$ independent realizations. For the hysteresis measurement, $\eta$ is swept quasi-statically, with the final configuration at each step used as the initial condition for the next, in both forward and backward directions. Fluctuations are characterized by the susceptibility $\chi=N(\langle v_a^2\rangle-\langle v_a\rangle^2)$ and Binder cumulant $G=1-\langle v_a^4\rangle/(3\langle v_a^2\rangle^2)$, where angular brackets denote averages over time and realizations. For a discontinuous transition $G$ develops a negative minimum, while for a continuous one it remains non-negative.

\begin{figure}[t]
        \centering
        \includegraphics[width=1\linewidth]{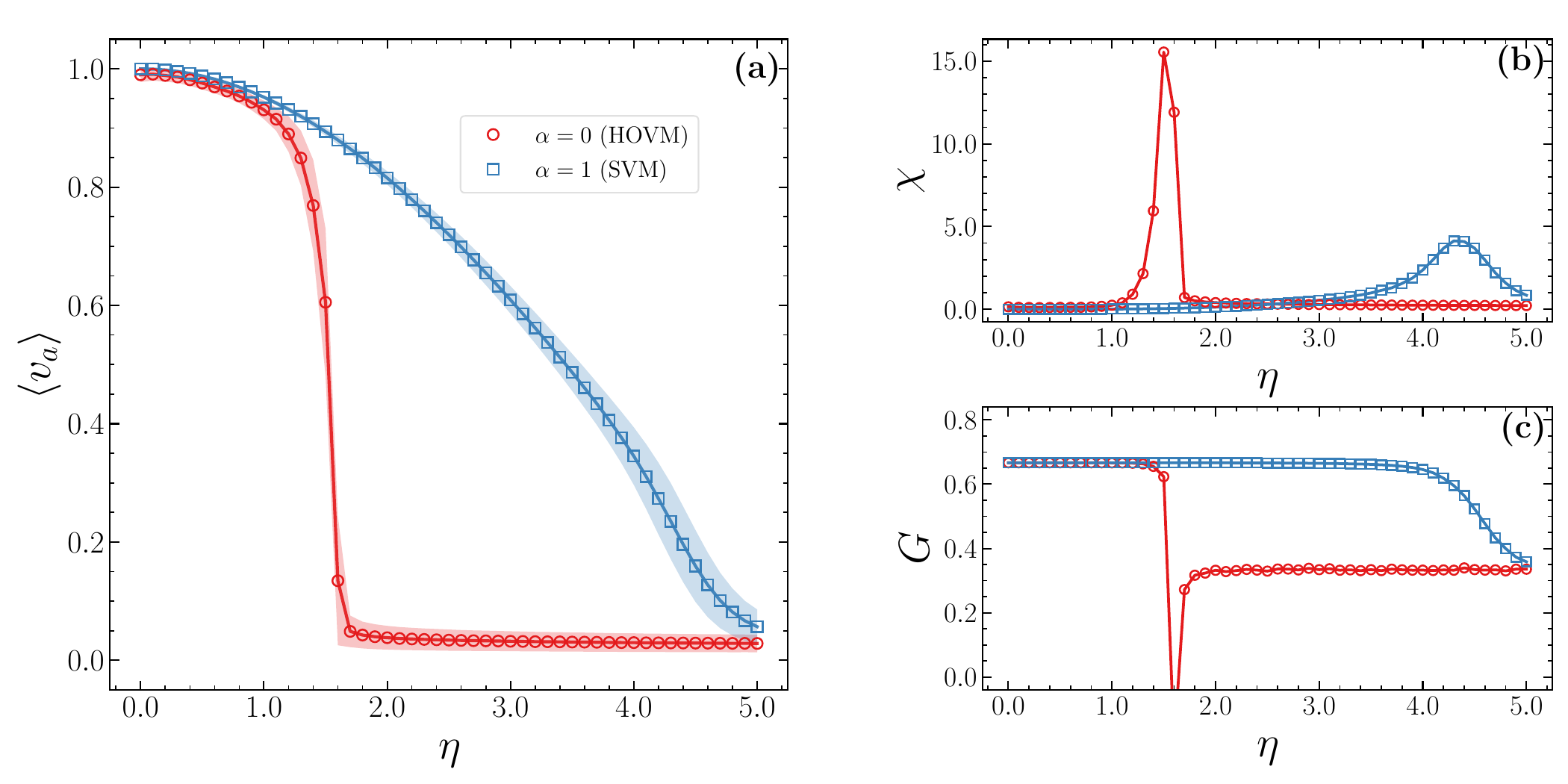}
        \caption{Order--disorder transition for $\alpha=1$ (SVM) and $\alpha=0$ (HOVM) at $\rho=8$ ($N=10^3$, $L=\sqrt{125}$). (a) Mean order parameter $\langle v_a\rangle$ versus noise $\eta$; shaded regions indicate statistical uncertainty from $10^3$ independent realizations. (b) Susceptibility $\chi$. (c) Binder cumulant $G$; the negative minimum for $\alpha=0$ is consistent with phase coexistence and a discontinuous transition.}
        \label{fig:main-fig2}
    \end{figure}

    The two limiting cases differ qualitatively in their transition character [Fig.~\ref{fig:main-fig2}]. In both, the order parameter decreases with increasing noise, but the decay is smooth for $\alpha=1$ (SVM) and abrupt for $\alpha=0$ (HOVM). The susceptibility shows a broad maximum for the SVM and a narrower peak for the HOVM, while the Binder cumulant remains non-negative for $\alpha=1$ but develops a negative dip for $\alpha=0$, consistent with phase coexistence.

\begin{figure}[t]
        \centering
        \includegraphics[width=1\linewidth]{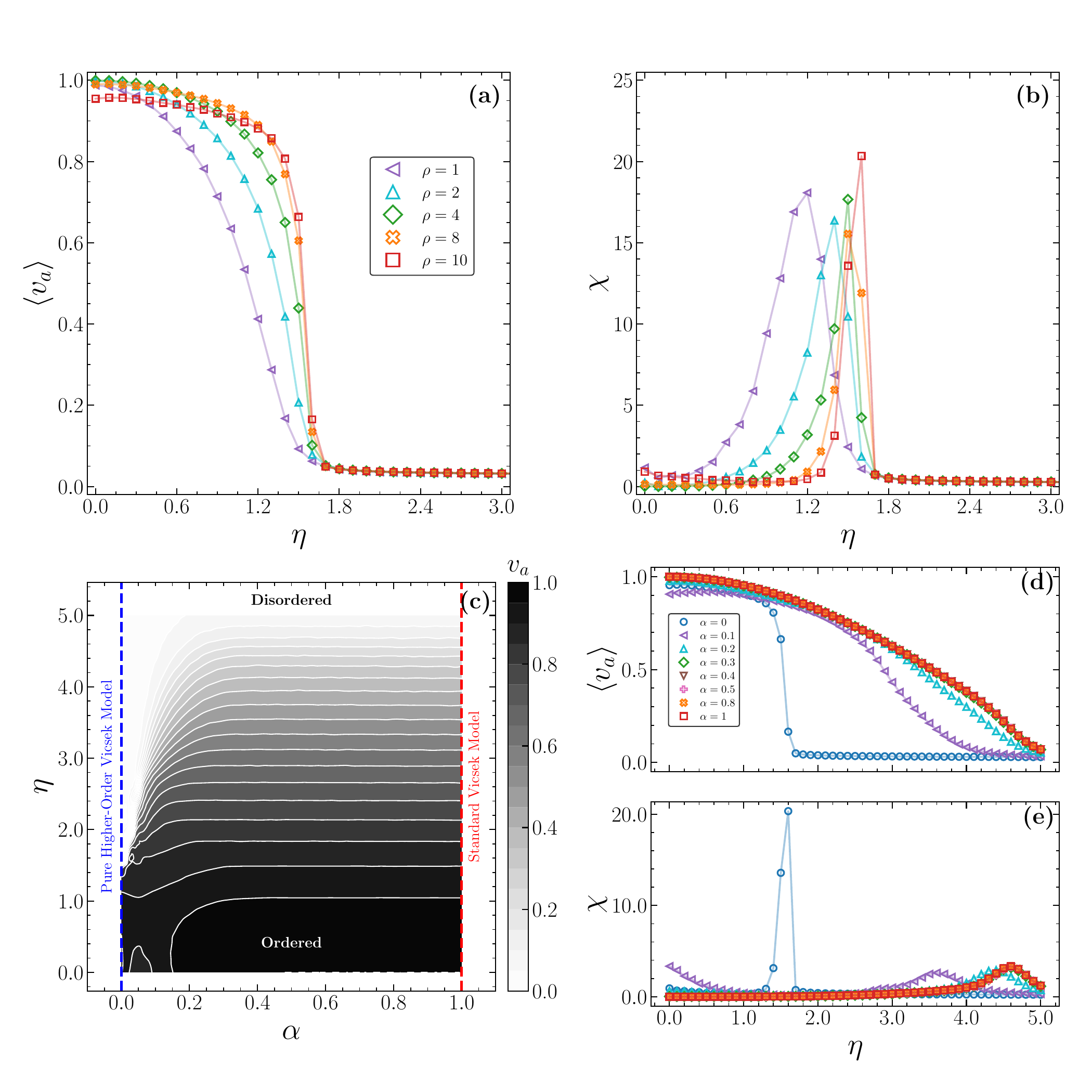}
        \caption{Density and interaction-mixing ($\alpha$) dependence of the order-disorder transition, averaged over $10^3$ realizations. (a),\,(b)~Fixed $\alpha=0$ (pure HOVM), $N=10^3$; $L$ varied to tune $\rho$. (a)~$\langle v_a\rangle$ vs.\ $\eta$. (b)~$\chi$ vs.\ $\eta$; peaks sharpen and shift to larger $\eta$ with increasing $\rho$, indicating that higher density stabilizes collective motion. (c)--(e)~Fixed $\rho=10$ ($N=10^3$, $L=10$), $\alpha$ varied from the pure HOVM ($\alpha=0$) to the SVM ($\alpha=1$). (c)~Contour map of $\langle v_a\rangle$ in the $(\alpha,\eta)$ plane; color encodes $\langle v_a\rangle$ from 0 (disordered) to 1 (ordered). (d)~$\langle v_a\rangle$ vs.\ $\eta$ for selected $\alpha$. (e)~$\chi$ vs.\ $\eta$; the peak shifts to larger $\eta$ and broadens with increasing $\alpha$, demonstrating that standard Vicsek alignment progressively extends the ordered phase.}
        \label{fig:main-fig3}
    \end{figure}

    The sharpening of the transition is systematic across both density and interaction type [Fig.~\ref{fig:main-fig3}]. At fixed $\alpha=0$, increasing $\rho$ shifts the transition to larger noise values and narrows the susceptibility peak, with the crowded regime ($\rho\ge8$) producing the sharpest behavior. In the $(\alpha,\eta)$ plane, the transition boundary sharpens progressively as $\alpha\to0$, confirming that higher-order interactions drive the crossover toward discontinuous behavior.

    \begin{figure}[t]
        \centering
        \includegraphics[width=1\linewidth]{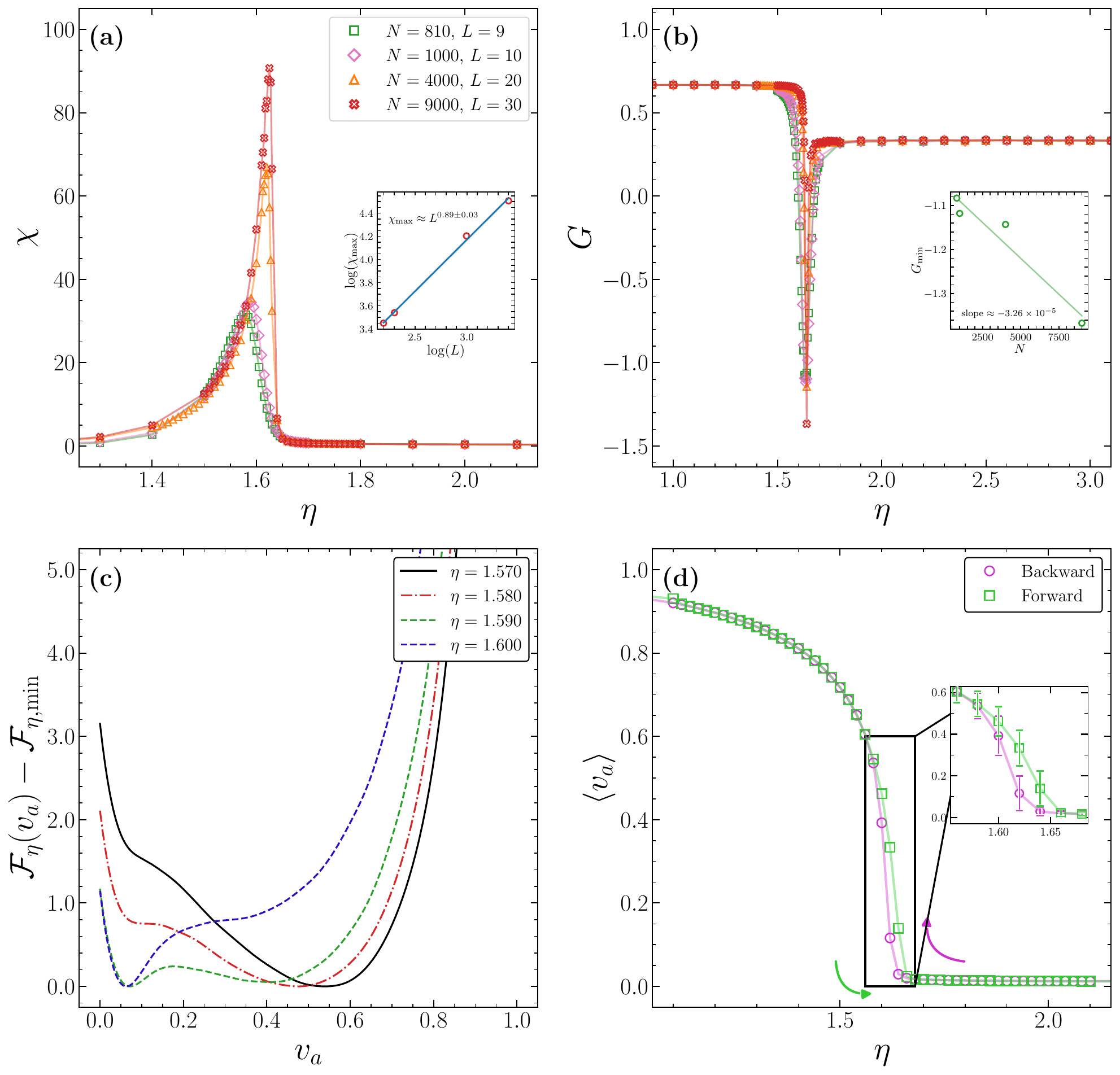}
        \caption{Finite-size scaling and discontinuous-transition signatures 
        at $\rho=10$, $\alpha=0$, averaged over $10^3$ independent realizations. (a)~Susceptibility $\chi$ vs.\ $\eta$ for different $N$ and $L$ at fixed $\rho=N/L^2$; the peak sharpens and grows with system size. inset: $\chi_{\max} \sim L^{\beta}$ on a log--log scale; a linear fit gives $\beta = 0.89 \pm 0.03$, indicating that susceptibility grows with system size near the transition. (b)~Binder cumulant $G$ vs.\ $\eta$; the negative minimum deepens and sharpens with increasing $N$, consistent with phase coexistence. Inset: $G_{\min}$ vs.\ $N$; the linear fit (slope $\approx -3.26\times10^{-5}$) confirms that $G_{\min}$ grows more negative with system size. (c)~Effective free energy $\mathcal{F}_{\eta}(v_a) = -\log\mathcal{P}_{\eta}(v_a) - \mathcal{F}_{\eta,\min}$ computed from the order-parameter distribution for $N=10^3$, $L=10$; at the critical noise $\eta_c=1.590$ a double-well structure emerges, signaling coexistence between the ordered and disordered phases, while noise values above and below $\eta_c$ show how the free energy deforms from a double-well to a single-well structure. (d)~Mean order parameter $\langle v_a\rangle$ under quasi-static forward (increasing $\eta$) and backward (decreasing $\eta$) noise sweeps for $N=9000$; inset: zoom of the hysteresis region with error bars.}
        \label{fig:main-fig4}
    \end{figure}

    \begin{figure}[h]
        \centering
        \includegraphics[width=1\linewidth]{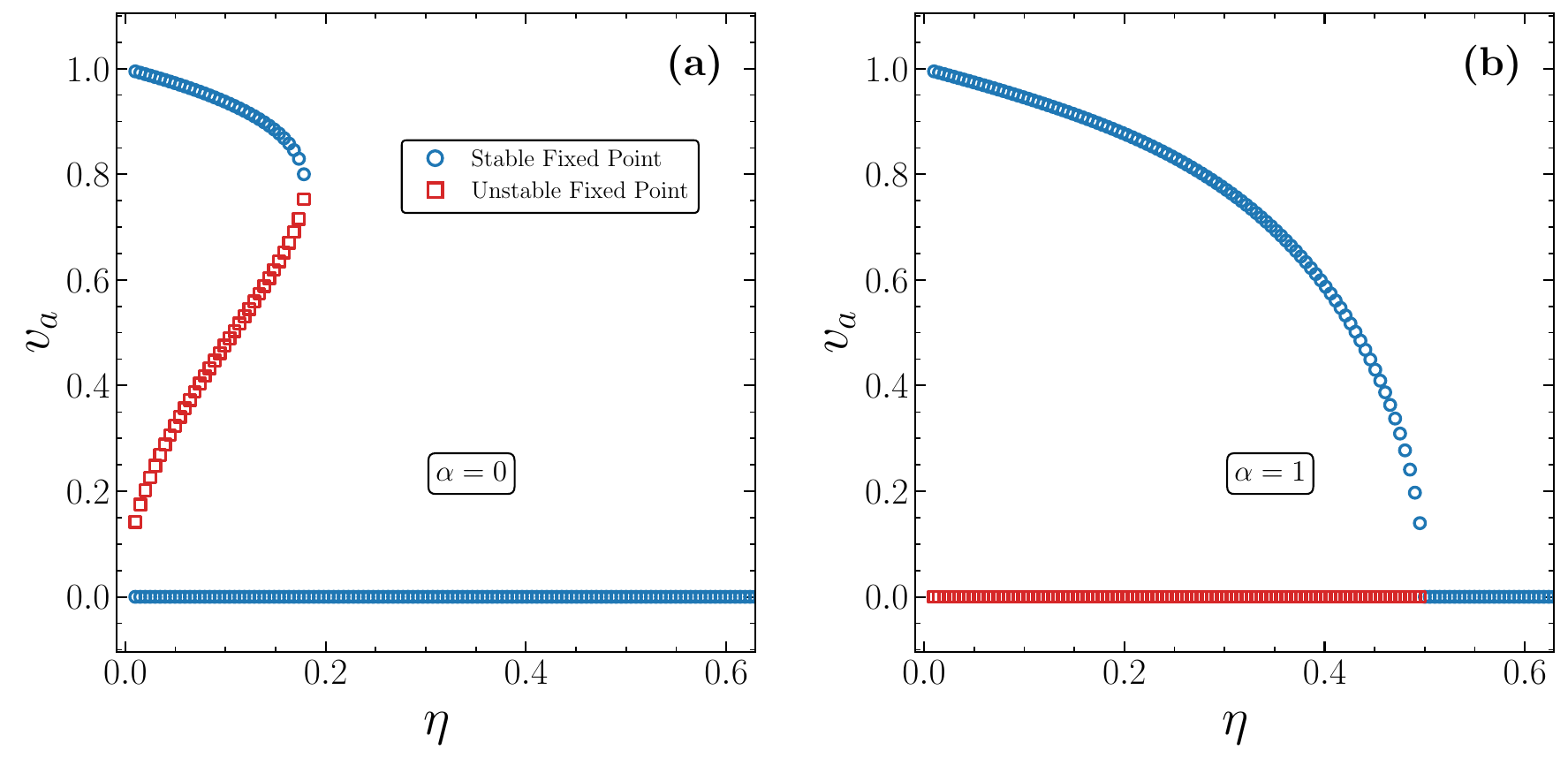}
        \caption{Mean-field fixed points of self-consistency condition for $\alpha=0$ (HOVM) and $\alpha=1$ (SVM). Open circles: stable fixed points; open squares: unstable fixed points. (a) $\alpha=0$: two stable branches separated by an unstable one, indicating bistability consistent with a discontinuous transition. (b) $\alpha=1$: a single stable branch vanishing continuously at $\eta_c$, consistent with a continuous transition. Note that the MF $\eta$ axis is not directly comparable to simulation values (see Sec.~\ref{sec:MF}).}
        \label{fig:main-fig5}
    \end{figure}
    
    Figure~\ref{fig:main-fig4} presents four independent signatures of a discontinuous transition at $\rho=10$, $\alpha=0$. The susceptibility $\chi$ develops a sharp peak that grows and narrows with system size [Fig.~\ref{fig:main-fig4}(a)], with $\chi_{\max}\sim L^{0.89\pm 0.03}$ (inset), indicating that fluctuations grow with system size near the transition. The Binder cumulant $G$ develops a negative minimum that deepens systematically with $N$ [Fig.~\ref{fig:main-fig4}(b)], a hallmark of phase coexistence absent in continuous transitions. The effective free energy $\mathcal{F}_{\eta}(v_a)=-\log\mathcal{P}_{\eta} (v_a)-\mathcal{F}_{\eta,\min}$, computed from the order-parameter 
    distribution at fixed $\eta$, develops a double-well structure near the transition for $N=10^3$ [Fig.~\ref{fig:main-fig4}(c)], directly signaling coexistence between the ordered and disordered phases. Finally, a quasi-static sweep of $\eta$ produces distinct forward and backward branches of $\langle v_a\rangle$ [Fig.~\ref{fig:main-fig4}(d)], confirming metastability and hysteresis characteristic of a first-order transition. Taken together, these four signatures provide consistent evidence for a discontinuous transition in the pure higher-order limit, which we interpret through a mean-field analysis in the following section. The simulation data, source code, and plotting scripts used to produce all figures in this work are publicly available on Zenodo~\cite{maryam_dataset}.

    \section{Order of the transition: mean-field mechanism}\label{sec:MF}
    To identify the mechanism driving the change in transition character, we apply a mean-field (MF) approximation in which spatial correlations are neglected and every particle experiences the same effective alignment field. Writing the global polarization as $\mathbf{P} = (Nv_0)^{-1}\sum_j \mathbf{v}_j = v_a(\cos\Theta,\sin\Theta)$, we replace all neighbor velocities by $\mathbf{v}_j \to v_a(\cos\Theta,\sin\Theta)$. Under this substitution, the pairwise field gives $\mathbf{p}_i \to v_a(\cos\Theta,\sin\Theta)$, contributing a term linear in $v_a$. For the higher-order field, each summand in Eq.~\eqref{eq:triadic_field} becomes $(\mathbf{v}_j\cdot\mathbf{v}_i)\mathbf{v}_k + (\mathbf{v}_k\cdot\mathbf{v}_i)\mathbf{v}_j \;\to\; 2v_a^2\cos(\theta_i-\Theta)\,v_0(\cos\Theta,\sin\Theta)$,where $\cos(\theta_i-\Theta)$ is the projection of particle $i$'s orientation onto the collective direction. Requiring self-consistency, $\langle\cos(\theta_i-\Theta)\rangle = v_a$, the higher-order contribution reduces to a term $\propto v_a^3$. Combining both terms, the effective alignment strength is $F(v_a) = \alpha\, v_a + (1-\alpha)\,v_a^3$. Assuming each particle's orientation is independently drawn from a von Mises distribution~\cite{mardiaBook}, $P(\theta)\propto\exp[\kappa\cos(\theta-\Theta)]$ with $\kappa = F(v_a)/\eta$, the self-consistency condition is $v_a = I_1(\kappa)/I_0(\kappa) \equiv f(v_a)$, whose fixed points $v_a^*$ are stable when $|f'(v_a^*)|<1$ (Fig.~\ref{fig:main-fig5}). For $\alpha=1$, a single branch vanishes continuously at $\eta_c$, consistent with a continuous transition. For $\alpha=0$, two stable branches separated by an unstable one emerge, reflecting bistability driven by the cubic term in effective alignment strength.
    
    This MF treatment is phenomenological: it neglects spatial correlations, density waves, and the inherently non-equilibrium character of the dynamics, which shift the transition point and cause a significant quantitative discrepancy with the simulations. The MF framework nevertheless isolates the essential mechanism: the cubic nonlinearity in $F(v_a)$ generates bistability in $v_a$, explaining why signatures of discontinuity emerge in the pure higher-order limit ($\alpha=0$) already at moderate system sizes.
 
    \section{Conclusion}
    Our study establishes that collective behavior can indeed emerge from higher-order interactions, and that such behavior is qualitatively distinct from its pairwise counterpart. Notably, the transition in the pure triadic limit occurs at lower noise amplitudes, implying that collective order sustained by higher-order rules is more sensitive to environmental fluctuations yet exhibits a discontinues phase transition with abrupt dynamical response.  A mean-field analysis identifies the mechanism: the triadic interaction introduces a nonlinearity in the effective alignment strength that generates bistability in the order parameter, absent in the pairwise case. This fragility, coupled with the sharp transition, suggests that systems governed by multi-agent interactions may be particularly well-suited for applications requiring rapid, coordinated decision-making under uncertainty, such as autonomous robotic swarms or distributed sensor networks capable of executing complex cooperative tasks.
      
    Higher-order interactions are well established in social, biological, and technological networks~\cite{Battiston2021, BOCCALETTI2023} as drivers of abrupt transitions and multistability absent in pairwise models, yet their role in self-propelled particle systems has remained largely unexplored. Our results establish a direct connection between these two fields, demonstrating that interaction structure controls the order of the phase transition independently of noise and density. These findings open a direction for investigating higher-order alignment rules in active matter, and biological collective motion.
    
	\bibliographystyle{apsrev4-2}
	\bibliography{HOVM_Ref}
	
	\appendix
	\renewcommand{\theequation}{S\arabic{equation}}
	\setcounter{equation}{0}
	\renewcommand{\thefigure}{S\arabic{figure}}
	\setcounter{figure}{0}

	\section{Decorrelation time and simulation protocol}\label{app:decorrelation}
	
	Successive configurations in time-dependent many-body simulations are temporally correlated and cannot be treated as independent samples. Reliable measurements therefore require an equilibration time long enough to remove memory of the initial condition and a sampling window long enough to contain effectively independent configurations. We determine these timescales from the autocorrelation of the global polarization vector,
	\begin{equation}
		\mathbf{P}(t) = \frac{1}{N v_0} \sum_{i=1}^{N} \mathbf{v}_i(t),
	\end{equation}
	where $\mathbf{v}_i(t)$ is the velocity of particle $i$, $N$ is the number of particles, and $v_0$ is the constant particle speed. We write $\mathbf{P}(t)=(P_x(t),P_y(t))$ and define its fluctuation around the stationary mean as
	\begin{equation}
		\delta \mathbf{P}(t) = \mathbf{P}(t) - \langle \mathbf{P} \rangle,
	\end{equation}
	where $\langle \mathbf{P} \rangle$ is evaluated after the system has reached the stationary state.
	
	The normalized autocorrelation function is
	\begin{equation}
		C(t) =
		\frac{
			\left\langle
			\delta \mathbf{P}(t_0)\cdot \delta \mathbf{P}(t_0 + t)
			\right\rangle_{t_0}
		}{
			\left\langle
			|\delta \mathbf{P}(t_0)|^2
			\right\rangle_{t_0}
		},
	\end{equation}
	where the average is taken over time origins $t_0$ along a stationary trajectory. This normalization gives $C(0)=1$, while the decay of $C(t)$ measures how fast the system loses memory of a previous configuration. The corresponding integrated autocorrelation time is
	\begin{equation}
		\tau = \int_0^{\infty} C(t)\, dt.
	\end{equation}
	In the simulations, $C(t)$ is computed from long stationary trajectories, and the integral is evaluated numerically up to the time where the autocorrelation has decayed to the level of statistical fluctuations. The integrated autocorrelation time gives the intrinsic decorrelation scale of the measured observable and is the standard quantity used to estimate independent sampling times in correlated many-body dynamics \cite{Sokal1997,ChatePRE2008}.
	
\begin{figure}[t]
		\centering
		\includegraphics[width=1\linewidth]{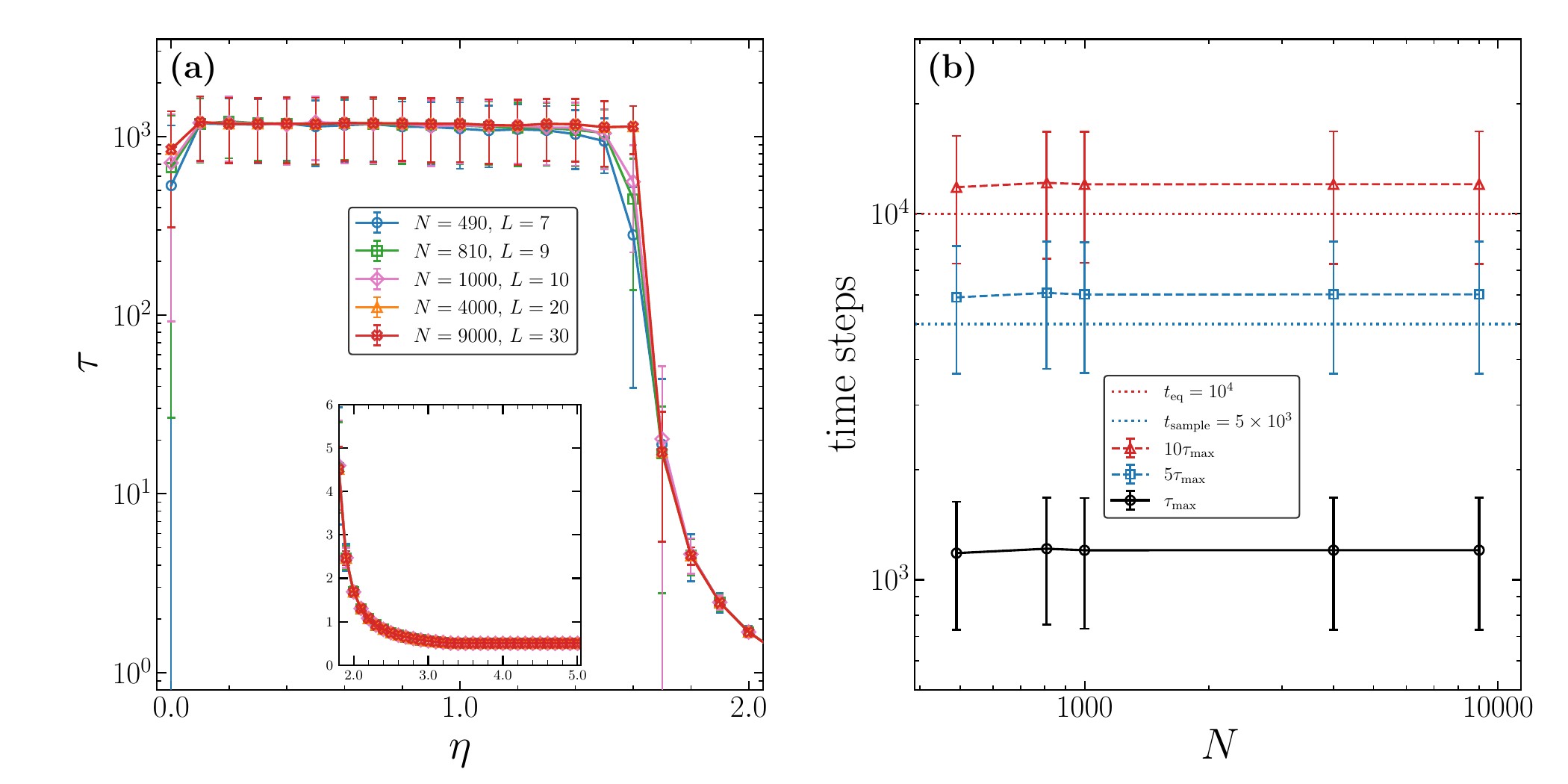}
		\caption{Decorrelation time analysis for $\alpha=0$ at $\rho=10$. (a) Integrated autocorrelation time $\tau$ versus noise amplitude $\eta$ for different system sizes. The inset shows the short-$\tau$ regime on a linear scale. (b) Maximum decorrelation time $\tau_{\max}$ versus system size, together with $5\tau_{\max}$ and $10\tau_{\max}$. Horizontal lines indicate the fixed simulation parameters $t_{\mathrm{s}}=5\times10^{3}$ and $t_{\mathrm{eq}}=10^{4}$. All quantities are averaged over $10^3$ independent realizations.}
		\label{fig:decorrelation_data1}
	\end{figure}

	Figure~\ref{fig:decorrelation_data1}(a) shows the measured $\tau$ as a function of the noise amplitude $\eta$ for several system sizes at $\rho=10$ and $\alpha=0$. In the disordered regime at large $\eta$, the autocorrelation decays rapidly and $\tau$ remains of order unity. Near the transition region, the decay becomes slower and $\tau$ increases, reflecting the growth of collective correlations. For the largest systems considered, the maximum values of $\tau$ are of order $10^3$. The inset shows the small-$\tau$ regime on a linear scale, making clear that the correlation time remains short away from the transition.
	
	We use the largest measured decorrelation time, denoted by $\tau_{\max}$, to choose conservative simulation times. Figure~\ref{fig:decorrelation_data1}(b) shows $\tau_{\max}$ as a function of system size, together with $5\tau_{\max}$ and $10\tau_{\max}$. The horizontal lines mark the fixed equilibration time and sampling window used in the simulations,
	\begin{equation}
		t_{\mathrm{eq}} = 10^4,
		\qquad
		t_{\mathrm{s}} = 5\times 10^3 .
	\end{equation}
	These values are chosen to satisfy
	\begin{equation}
		t_{\mathrm{eq}} \gtrsim 10\,\tau_{\max},
	\end{equation}
	and
	\begin{equation}
		t_{\mathrm{s}} \gtrsim 5\,\tau_{\max}.
	\end{equation}
	Thus, the equilibration stage is long compared with the slowest measured decorrelation time, and the sampling window is long enough to contain several effectively independent configurations.
	
\begin{figure}[H]
		\centering
		\includegraphics[width=1\linewidth]{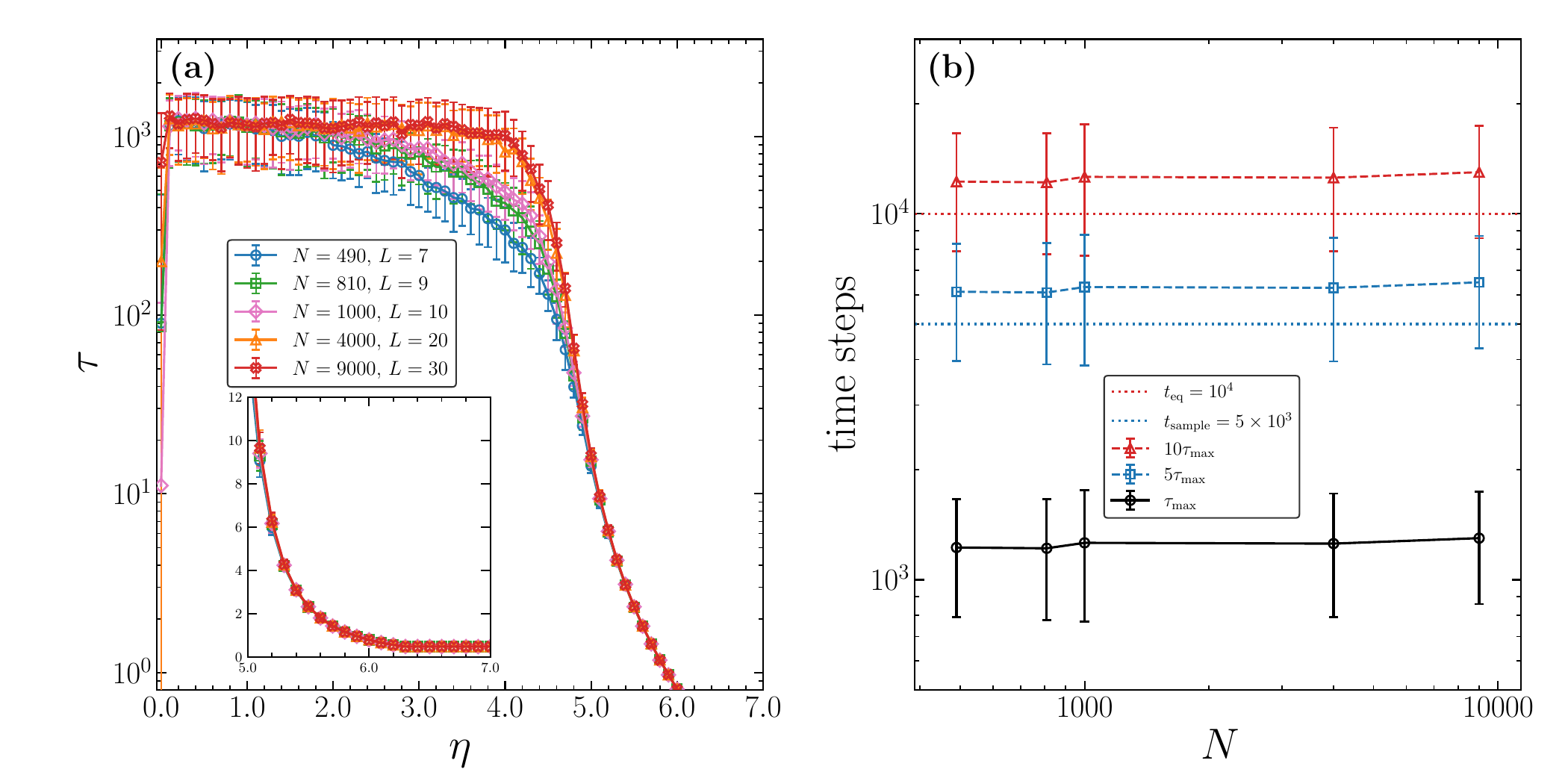}
		\caption{Decorrelation time analysis for $\alpha=1$ at $\rho=10$. (a) Integrated autocorrelation time $\tau$ versus noise amplitude $\eta$ for different system sizes. The inset shows the short-$\tau$ regime on a linear scale. (b) Maximum decorrelation time $\tau_{\max}$ versus system size, together with $5\tau_{\max}$ and $10\tau_{\max}$. Horizontal lines indicate the fixed simulation parameters $t_{\mathrm{s}}=5\times10^{3}$ and $t_{\mathrm{eq}}=10^{4}$. All quantities are averaged over $10^3$ independent realizations.}
		\label{fig:decorrelation_data2}
	\end{figure}

	Each simulation is first evolved for $t_{\mathrm{eq}}$ time steps without recording observables. Measurements are then collected over the sampling window $t_{\mathrm{s}}$, and averages are taken over both time and independent realizations. The data in Fig.~\ref{fig:decorrelation_data2} are averaged over $10^3$ independent realizations. We also checked that the decorrelation times for $\alpha=1$ and representative intermediate values of $\alpha$ are of the same order as those shown for $\alpha=0$. The same choices of $t_{\mathrm{eq}}$ and $t_{\mathrm{s}}$ therefore provide equilibrated and effectively decorrelated samples throughout the full range $\alpha \in [0,1]$.
	
\end{document}